\documentclass[aps, twocolumn, showpacs,superscriptaddress]{revtex4-1}

\usepackage{amssymb}
\usepackage{amsmath}
\usepackage[T2A]{fontenc}
\usepackage{graphics}
\usepackage{graphicx}
\usepackage[dvips]{color}

\begin{document}

\title{Kinetics of electron cooling in metal films at low temperatures  and revision of the two-temperature
model}

\author{A. I. Bezuglyj}\email{bezuglyj@kipt.kharkov.ua}
\affiliation{\it National Science Center "Kharkov Institute of Physics and Technology", 1,
Akademicheskaya St., Kharkov, 61108, Ukraine\\} \affiliation{\it Kharkov National University, 4,
Svobody Sq., Kharkov, 61077, Ukraine }

\author{V. A. Shklovskij}\email{shklovskij@univer.kharkov.ua}
\affiliation{\it Kharkov National University, 4, Svobody Sq., Kharkov, 61077, Ukraine }

\date{\today}

\begin{abstract}

The two-temperature model (2TM) introduced by Kaganov, Lifshitz, and Tanatarov is widely used to
describe the energy relaxation in the electron-phonon system of a metallic film. At the same time,
 the accuracy of the description of the  electron-phonon system in terms of 2TM,
i.e. on the basis of the electron and phonon temperatures, has not been considered in detail until
now. In this paper we present a microscopic theory of cooling of instantly heated electrons in
metallic films. In  framework of this theory the main features of electron cooling in thick and
thin films were found, and an analysis of the accuracy of the 2TM in the low-temperature region
was carried out. We consider a more accurate three-temperature model, which (in contrast to 2TM)
explicitly takes into account phonons with angles of incidence exceeding the angle of total
internal reflection. The contribution of these phonons to the kinetics of electron relaxation can
be  significant if the sound velocities in the film and the substrate are quite different.

\end{abstract}


\maketitle

\section{Introduction}

The two-temperature model proposed by Kaganov, Lifshitz, and Tanatarov in 1956
\cite{KaganovLifshitzTanatarov} is still the main model for analyzing experiments on the energy
relaxation of excited electrons in metals. It is sufficient to notice that on the basis of this
model, the effect of high-power laser radiation on the metal surface was considered
\cite{Anisimov1970, Anisimov1974}, and the kinetics of electron cooling in the experiments on the
heating of metal films by femtosecond laser pulses  was analyzed \cite{Fujimoto1984,
Elsayed-Ali1987, Schoenlien1987}. The essence of 2TM is as follows. It is believed that excited
electrons lose their energy as a result of the emission of phonons. In the process, both the
electron and phonon subsystems are assumed to be thermalized, i.e. having temperatures $ T_e $ and
$ T_p $, respectively. The energy transfer from electrons to phonons, $ P_ {ep} $, is expressed in
terms of the phonon-electron collision integral, in which the electron distribution function is a
Fermi function with temperature $ T_e $, and the phonon distribution function is Bose function
with temperature $ T_p $. In the deformation potential approximation, a direct calculation yields
the relation $ P_{ep} = f (T_e) - f (T_p) $ with the function $ f (T) = GT $ at temperatures much
higher the Debye temperature $ \Theta_D $, and $ f ( T) = \Sigma T^5 $ for $ T \ll \Theta_D $. The
constants $ G $ and $ \Sigma $, which depend on the material, characterize the strength of the
electron-phonon interaction at the high and low temperature ranges.

In the work of one of the authors \cite{ShklovskiiJLTP1980}, the two-temperature approach was
generalized to the case of an arbitrary electron dispersion law and an arbitrary frequency
dependence of the Eliashberg function $ \alpha^{2} F (\omega) $, describing the interaction of
electrons and phonons. With the help of a similar generalization, Allen \cite{AllenPRL1987}
deduced the linear relation between the cooling rate of a thin film and the electron-phonon
coupling constant $ \lambda $ which is well known in the theory of superconductivity. The relation
established by Allen made it possible to obtain the values of $ \lambda $ for a number of metals
and compounds from the experiment (see, for example, Ref.~\onlinecite{Brorson1990}).

Despite the progress in understanding the peculiarities of the relaxation of heated electrons in
metals, the validity and the range of the applicability of 2TM have not yet been fully elucidated
\cite{FannPRL1992, FannPR1992, GroeneveldPR1992, GroeneveldPR1995, AhnPR2004} and require a
special analysis especially in the low-temperature region, where 2TM is used to describe the work
of hot electron bolometers. \cite{Semenov-Nebosis, Nebosis} A  microscopic theory based on kinetic
equations for the electron and phonon distribution functions can provide reliability of such an
analysis. A kinetic approach to the analysis of steady-state heat transfer between a metal layer
and insulator slabs was developed in  Refs.~\onlinecite{ShklovskiiJLTP1980, ShklovskiiJETP1980}.
Subsequently, this approach was extended to the case of unsteady heating of a film deposited on a
dielectric substrate \cite{BezuglyjShklovskijJETP1997} (see also
Ref.~\onlinecite{Bezuglyj&Shklovskij2014}). The equations obtained in
Ref.~\onlinecite{BezuglyjShklovskijJETP1997} will be used below to compare the predictions of 2TM
with the results of a microscopic model. This will make it possible to determine whether the model
based on the assumption of phonon and electron temperatures
 adequately describes the energy relaxation of electrons. Thus, the purpose of this paper is
 twofold: firstly, to
obtain a fairly complete physical picture of the energy relaxation of heated electrons from the
microscopic theory, and secondly, to find out to what degree the  predictions of 2TM correspond to
this picture. The analysis is limited to a region of low temperatures $ T <T_*$, where
electron-electron collisions dominate electron-phonon collisions, which allows us to consider
electrons as being thermalized. Because the temperature $ T_* \sim \Theta_{D}^2 / \varepsilon_F $,
for typical metals we have $ T_* \sim 1 \, $ K. It should also be noted that in sufficiently
"dirty" metals the temperature $ T_* \lesssim 10 \, $ K due to the enhancement of the
electron-electron and the weakening of the electron-phonon interaction. \cite{GershenzonJETPL1982}

Unlike electrons, phonons are not thermalized at low temperatures, since at $ T \ll \Theta_{D} $
the frequency of phonon-phonon collisions is several orders of magnitude lower than the collision
frequency of phonons with electrons. This condition is  sufficient to avoid introducing the phonon
temperature and to describe the phonon subsystem in terms of the distribution function. The exact
solution of the kinetic equation for the phonon distribution function obtained in
Ref.~\onlinecite{BezuglyjShklovskijJETP1997} allows us to represent the heat balance equation for
electrons in the form of a nonlinear integro-differential dynamic equation for the electron
temperature. This equation is the basis for analyzing the cooling of instantly heated electrons in
metallic films at low temperatures. In this paper we consider the case of relatively weak heating,
when the equation can be solved using the operational method (the Laplace transformation).
Comparison of Laplace transforms of the electron temperature, found from microscopic theory and
2TM, allows us to establish the range of applicability of 2TM and to analyze its accuracy.

The energy relaxation is described in 2TM by two linear differential equations - for the electron
and the phonon temperatures. However, 2TM does not take into account the fact that the phonons are
divided into two groups by their ability to pass through the interface between the film and the
substrate. Indeed, if $ \theta_{cr} $ is the critical angle of total internal reflection, then
phonons with angles of incidence $ \theta <\theta_{cr} $ can pass to the substrate with some
probability $ \alpha (\theta) $, and phonons with $ \theta_{cr} <\theta <\pi / 2 $ are completely
reflected from the boundary with the substrate. The three-temperature model (3TM) assigns the
temperature $ T_e $ to the electrons, and  the temperatures $T_{p1} $ and $ T_{p2} $ to the
above-mentioned phonon groups, respectively. Below we show that  the 3TM significantly improves
the description of the relaxation of the electron temperature compared with 2TM and in a number of
cases it yields results practically not differing from the results of the microscopic model.
Notice that 3TM is formulated in accordance with the so-called "multi-temperature approach", when
each subsystem of particles participating in the energy relaxation is assigned its own temperature
(see, for example, Refs.~\onlinecite{MansartPRB2010, ParlatoPRB2013, Sidorova2016}).

The structure of this article is as follows. In Sec.~\ref {s2} for the convenience of the reader,
we reproduce the main points of the derivation of the dynamic equation for the electron
temperature \cite{BezuglyjShklovskijJETP1997}. In Sec.~\ref {s3} on the basis of this equation, we
analyze the dynamics of cooling of electrons in a linear regime, i.e. when the initial heating of
the electrons is small in comparison with the temperature of the thermostat. Here, the case of a
heat-insulated film is considered. The heat transfer to the substrate is taken into account in
Sec.~\ref {s4}, which presents the analytical results describing the cooling of electrons in pure
films. In the same Section, a three-temperature model is considered and a comparison of the
predictions of the microscopic model, 2TM, and 3TM is made. Sec.~\ref {s5} contains a discussion
of the results of the work and the main conclusions.

\section{Microscopic derivation of the dynamic equation for the electron temperature}\label{s2}

In order to examine the features of the relaxation dynamics of the electron temperature $ T_e $,
we use a simple microscopic model that allows obtaining the main results in an analytical form.
So, let's take the spectrum of electrons as quadratic and isotropic, $ \epsilon_{\bf p} = p^2 /2m
$ ($ {\bf p} $ is the quasimomentum of the electron, $ m $ is its effective mass) and suppose that
phonons have a longitudinal acoustic branch only. Since we confine ourselves to a temperature
region well below the Debye temperature $ \Theta_D $, the dispersion law for phonons can be
considered linear: $ \omega_{\bf q} = sq $, where $ s $ is the velocity of longitudinal sound, $ q
$ is the modulus of the phonon wave  vector. For the electron-phonon interaction, we use the
deformation potential approximation, modified in such a way as to take into account the
renormalization of this interaction by impurities. Finally, the heat exchange between the metal
film and the dielectric substrate will be described in terms of the well-known acoustic mismatch
model \cite{Little, SwartzPohl, Kaplan}.

In the low-temperature region, electron-electron collisions dominate over electron-phonon
collisions, and this means that the electronic subsystem of the film is thermalized. In the
absence of a transport current,  the electron distribution function is the isotropic Fermi
function. Due to the high thermal conductivity of the electron gas, the electron temperature $ T_e
$ does not depend on the coordinate $ z $ perpendicular to the film, but depends only on the time
$ t $ \cite{ShklovskiiJETP1980}. The phonon distribution function $ N_{\bf q} (z, t) $ must be
found from the kinetic equation

\begin{equation}\label{1}
{{\partial N_{\bf q}}\over{\partial t}} + s_z{{\partial N_{\bf q}} \over{\partial z}} = \nu_q
\{n_q [T_{e}(t)] - N_{\bf q}(z,t) \},
\end{equation}
where $ s_z $ is the projection of the phonon velocity on the axis $ z $, $ \nu_q $ is the
phonon-electron collision frequency, $ n_q (T) = [\exp (\hbar \omega_q / T) -1]^{-1} $ is the Bose
distribution function ($ k_B = 1 $). A fairly simple form of the integral of phonon-electron
collisions (the right-hand side of the kinetic equation) is a consequence of the Fermi
distribution of electrons. For pure films, in the deformation potential approximation, the
frequency of phonon-electron collisions is given by

\begin{equation}\label{2}
\nu_q={{m^2 \mu^{2}} \over{2\pi \hbar^3 \rho_f s}}\omega_q.
\end{equation}
Here $ \mu $ is the deformation potential constant, which is of the order of the Fermi energy
$\epsilon_{F} $; $ \rho_f $ is the film density. The linear dependence of $ \nu_q $ on $ \omega_q
$ corresponds to the quadratic dependence of the Eliashberg function on the phonon frequency, $
\alpha^{2}F(\omega) \propto \omega^2 $, and to the cubic dependence of the electron-phonon
collision frequency on temperature: $ \tau_{e}^{- 1} \propto T_{e}^{3} $. \cite{RoukesPRL1985,
WellstoodPRB1994, YungAPL2002, VinantePRB2007}

In dirty films, where the electron mean free path $ l $ is much smaller than the wavelength of the
thermal phonons $ \lambda_T $ ($ \lambda_T = 2 \pi \hbar s / T_e) $, the dependencies as $
\tau_{e}^{-1} \propto T_{e}^{4} $ \cite{GershensonAPL2001, KarvonenPR2005, KarvonenJLTP2007,
LinJPhysCondMatt2002} and $ \tau_ {e}^{- 1} \propto T_{e}^{2} $ \cite{LinJPhysCondMatt2002,
KarasikIEEEApplSupercond2007, WeiNatureNanotechnol2008, GershensonJETP1990, BergmannPRB1990} were
observed experimentally. In order to extend our analysis to the case of dirty films, we generalize
Eq.~(\ref{2}) and write down the frequency of phonon-electron collisions in the form

\begin{equation}\label{3}
\nu_q={{m^2 \mu_{r}^{2}} \over{2\pi \hbar^4 \rho_f s}} (\hbar\omega_q)^{1+r}.
\end{equation}
where $ \mu_r $ is a phenomenological parameter, and $ r $ is a number. Equation (\ref {3}) leads
to a power-law frequency dependence of the Eliashberg function: $ \alpha^2F(\omega) \propto
\omega^{2 + r} $ \cite{BergmannPRB1990} and to the power-law temperature dependence of the
electron-phonon frequency collisions of $ \tau_{e}^{- 1} \propto T_{e}^{3 + r} $. So, Eq.~(\ref
{3}) describes both groups of experiments on dirty films, if we take $ r = 1 $ and $ r = -1 $,
respectively. As follows from the theory \cite{SergeevPRB2000}, the dependence $\tau_{e}^{- 1}
\propto T_{e}^{4} $ occurs due to the scattering of electrons by light point defects that
oscillate together with the crystal lattice, whereas $ \tau_ {e}^{- 1} \propto T_{e}^{2} $
corresponds to the dominant scattering of electrons by heavy impurity atoms or grain boundaries,
which are weakly  involved in lattice vibrations.

Boundary conditions should be added to the differential equation (\ref {1}). As one of them we
take the condition of specular reflection of phonons on the free boundary of the film (at $z=d$):

\begin{equation}\label{4}
N_{\bf q}(d,t) = N_{\bf q^{\prime}}(d,t).
\end{equation}
Here $ {\bf q} = (q_x, q_y, q_z> 0) $ is the wave vector of the phonon incident on the boundary $
z = d $, and $ {\bf q ^ {\prime}} = (q_x, q_y, -q_z) $ is the wave vector of the specularly
reflected phonon. At the interface between the film and the substrate ($ z = 0 $), the following
condition is fulfilled:

\begin{equation}\label{5}
N_{\bf q}(0,t)=\alpha n_{q}(T_B)+\beta N_{\bf q^{\prime}}(0,t),
\end{equation}
where $ \alpha $ is the probability of the phonon passing the film/substrate interface, and $
\beta = 1- \alpha $ is the probability of the phonon reflection from the interface. The condition
(\ref {5}) means that both phonons reflected from the interface with the substrate and phonons
 that have passed from the substrate to the film  fall into the state with the wave vector
 $ {\bf q} $; $ T_B $ is the temperature of the substrate. Note that the condition (\ref {5}) assumes
that phonons that leave the film do not return back. Such a picture corresponds to  experiments on
fairly narrow films deposited on single-crystal substrates with high thermal conductivity.

\begin{figure}[t]
\includegraphics[width=9.5 cm]{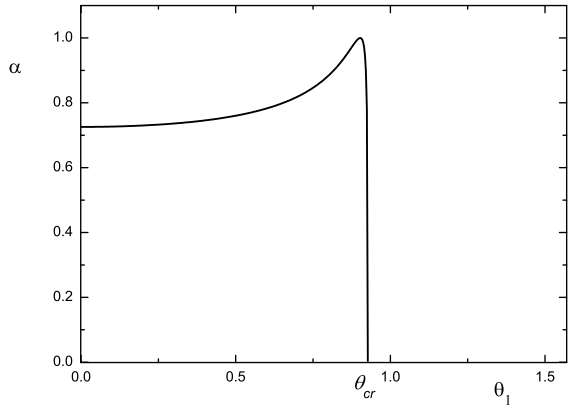}
\caption{\label{f1}The probability of the passage of phonons through the Cu/SiO$_2 $ interface.
The following parameter values were used to calculate $ \alpha (\theta_1) $: $ s_{Cu} $ = 4.7
km/s, $ s_{SiO_2} $ = 5.9 km/s, $ \rho_{Cu} $ = 8.9 g/cm$^3 $ and $ \rho_{SiO_2} $ = 2.2 g/cm$^3
$.}
\end{figure}

In the acoustic mismatch model \cite{Little, SwartzPohl, Kaplan} the probability $ \alpha $
depends on the angle of incidence of the phonon and the acoustic impedances of the film and
substrate:

\begin{equation}\label{6}
\alpha(\theta_1)= {\frac{4ZZ^{\prime}{\rm cos}\theta_1{\rm cos}\theta_2}{(Z{\rm cos}\theta_2 +
Z^{\prime}{\rm cos} \theta_1)^2}}.
\end{equation}
Here $ Z = \rho_f s $ ($ Z^{\prime} = \rho^{\prime} s^{\prime} $) is the acoustic impedance of the
film (substrate); the angles of incidence and refraction are related by the relation $ s^{\prime}
{\rm sin} \theta_1 = s \, {\rm sin} \theta_2 $  ($ \rho^ {\prime} $ and $ s^{\prime} $ denote the
density and velocity of the longitudinal sound of the substrate). The dependence of $ \alpha
(\theta_1) $ for a copper film on a quartz substrate is shown in  Fig.~\ref {f1}. The angle of
total internal reflection is $ \theta_{cr} = \arcsin s / s' $. Since $ s $ = 4.7 km/s, $ s' $ =
5.9 km /s, the angle $ \theta_{cr}$ is approximately   0.92.

The solution of the kinetic equation (\ref {1}) with the boundary conditions (\ref {4}) and (\ref
{5}) can be obtained by  performing the Fourier transformation with respect to time, and then
solving the ordinary differential equation for $ N_{ \bf q} (z, \omega) $ (see the details of
calculations in Ref.~\onlinecite{BezuglyjShklovskijJETP1997}). The inverse Fourier transformation
gives

\begin{eqnarray}\label{7}
N_{\bf q}(z,t) = \alpha[1-\beta \chi (0)]^{-1}\exp[-z\nu_q /s_z ]n_q(T_B) \nonumber \\
 + \int_{-\infty}^{t}dt^{\prime} n_q [T_e (t^{\prime})]
\nu_q \exp[-\nu_q (t-t^{\prime})]\beta^{[\tau +1 -z/2d]}\,,
\end{eqnarray}

\begin{eqnarray}\label{8}
&N&_{{\bf q}^{\prime}}(z,t) = \alpha[1-\beta \chi (0)]^{-1}\exp[-(2d-z)\nu_q /\vert s_z \vert]
n_q(T_B) \nonumber \\
 &+&\int_{-\infty}^{t}dt^{\prime} n_q [T_e (t^{\prime})]
\nu_q \exp[-\nu_q (t-t^{\prime})]\beta^{[\tau +z/2d]} ,
\end{eqnarray}
with $ \chi (0) = \exp{(-2d \nu_q / \vert s_z \vert)} $ and $ \tau = \vert s_z \vert
(t-t^{\prime}) / 2d $. The square brackets in the exponent $ \beta $ denote the integer part of
the number.

The equation for the electron temperature follows from the Boltzmann kinetic equation for the
electron distribution function. Multiplying this equation by the energy of the electron and
integrating over the momenta, we obtain  the heat balance equation

\begin{equation}\label{9}
c_e(T_e) {{dT_e}\over {dt}} = W(t) - P_{ep}(t),
\end{equation}
with the electronic heat capacity $ c_e = (\pi^2/3) N (0) T_e $, where $ N (0) = mp_F / \pi^2
\hbar^3 $ is the density of states on the Fermi surface. The symbol $ W (t) $ represents the
specific power of the heat sources in the film. The average specific power transferred from the
electrons to the phonons  is written as follows:

\begin{equation}\label{10}
P_{ep} = {1\over d} \int _0^d dz \int{{d^3 q}\over{(2\pi)^3}}\hbar\omega \nu_q \{n_q [T_{e}(t)] -
N_{\bf q}(z,t) \}.
\end{equation}
Since $ N_{\bf q} (z, t) $ is known, further calculations are not difficult. Substituting $P_{ep}$
in  Eq.~(\ref {9}), we obtain a nonlinear integro-differential equation for the electron
temperature

\begin{eqnarray}\label{11}
&c&_e(T_e) {{dT_e}\over {dt}} = W(t) - 2\int_{q_z > 0}{{d^3 q}\over{(2\pi)^3}}\hbar\omega_{q}
\nu_q \Bigl\{n_q (T_e (t))\nonumber \\
 &-& n_q(T_B) -\int_{-\infty}^{t}dt^{\prime} [n_q (T_e
(t^{\prime}))- n_q(T_B)]\nonumber \\
 &\times&\nu_q \exp[-\nu_q (t-t^{\prime})]\beta^{[\tau]}(1-\alpha
\{\tau\})\Bigr\},
\end{eqnarray}
where $ [\tau] $ and $ \{\tau \} $ denote the integer and fractional part of $ \tau $,
respectively. The appearance of these quantities is a consequence of  multiple reflections of
phonons from the boundaries of the film. We notice that the phonon contribution to the heat
balance equation consists of two parts. The first of them, local in time, describes the radiation
of nonequilibrium phonons at time $ t $, and the second part, integral with respect to time,
represents the absorption of nonequilibrium phonons emitted at earlier moments $ t^{\prime} <t $.

In what follows, we shall consider the kinetics of cooling of electrons heated during the period
much less than the characteristic times of thermal relaxation. If such heating ends at the time $
t = 0 $, then in Eq.~(\ref {11}) one can take $ W (t) = w \delta (t) $, where the absorbed energy
$ w $ determines the electron temperature jump from temperature of the thermostat $ T_B $ to
$T_e(0)$ according to the equality $ 2w = c_e (T_e (0)) T_e (0) -c_e (T_B) T_B $. After the
heating of the electronic subsystem, its energy relaxation begins. The dynamics of relaxation is
described by Eq.~(\ref {11}), where we should put $ W (t) = 0 $ and take the lower limit of time
integration equal to zero.

The integro-differential equation (\ref {11}) can be linearized in the case of weak heating, when
$ \Delta T_e (t) = T_e (t) -T_B \ll T_B $. The linearized equation, having the form

\begin{eqnarray}\label{12}
&c&_e(T_B) {{d\Delta T_e (t)}\over {dt}} = - 2\int_{q_z>0}{{d^3 q}\over{(2\pi)^3}}\hbar\omega_{q}
\nu_q
\frac{d n_q (T_B)}{d T_B}\Bigl\{ \Delta T_e (t) \nonumber \\
&-&\int_{0}^{t}dt^{\prime}\Delta T_e (t^{\prime}) \nu_q \exp[-\nu_q
(t-t^{\prime})]\beta^{[\tau]}(1-\alpha \{\tau\})\Bigr\},
\end{eqnarray}
is the basis for subsequent analysis.

We notice that the case of weak heating is quite meaningful, since it allows us to find all the
cooling regimes of electrons and their characteristic times, and also to compare the predictions
of the two-temperature and microscopic models. Moreover, it is possible to modify 2TM and
formulate a three-temperature model (3TM), which gives a picture of the energy relaxation of
electrons very close to the predictions of the microscopic theory.

\section{Relaxation of the electron temperature in a thermally insulated film}\label{s3}

First of all, let us consider the case when the electrons cool down in the absence of heat escape
from the film to the substrate. Since in reality the thermal insulation of the film from the
substrate cannot be ideal, in Sec.~\ref {s4} we establish the conditions under which the film can
be considered sufficiently insulated. Now we just put $ \alpha = 0 $ and $ \beta = 1 $ in
Eq.~(\ref {12}) and obtain

\begin{eqnarray}\label{13}
c_e(T_B) {{d\Delta T_e (t)}\over {dt}} =  - \int{{d^3 q}\over{(2\pi)^3}}\hbar\omega_{q} \nu_q
\frac{d n_q (T_B)}{d T_B}
\Bigl\{\Delta T_e (t) \nonumber \\
-\int_{0}^{t}dt^{\prime} \Delta T_e (t^{\prime}) \nu_q \exp[-\nu_q
(t-t^{\prime})]\Bigr\}.\,\,\,\,\,\,\,\,\,\,
\end{eqnarray}
The natural method for solving this equation is the operational method (the Laplace
transformation). For the Laplace transform of the electron temperature

\begin{equation}\label{14}
\Delta \tilde{T}_e (p) = \int_{0}^{\infty}\Delta T_e (t) e^{-pt}dt,
\end{equation}
the operational method gives

\begin{equation}\label{15}
\Delta \tilde{T}_e (p) = \frac{c_e \Delta T_e (0)}{c_e p + a -\tilde{K}(p)}.
\end{equation}
Here $ p $-independent quantity $ a $ is determined by the integral over the phonon wave  vectors
as

\begin{equation}\label{16}
a =  \int{{d^3 q}\over{(2\pi)^3}}\hbar\omega_{q} \nu_q \frac{d n_q (T_B)}{d T_B},
\end{equation}
and $ \tilde {K} (p) $ is the Laplace transform of the kernel of Eq.~(\ref {13}):

\begin{equation}\label{17}
\tilde{K}(p)=  \int{{d^3 q}\over{(2\pi)^3}}\hbar\omega_{q}  \frac{d n_q (T_B)}{d
T_B}\frac{\nu_{q}^{2}}{p+\nu_q}.
\end{equation}

Converting the Laplace transformation, we obtain the required time dependence of the electron
temperature in the form of an integral in the complex $ p-$plane

\begin{equation}\label{18}
\Delta T_e (t)= \frac{1}{2\pi i} \int_{\sigma-i\infty}^{\sigma+i\infty} \frac{e^{pt} c_e \Delta
T_e (0)\,dp} {p\Bigl[c_e+\int{{d^3 q}\over{(2\pi)^3}}\hbar\omega_{q} \frac{d n_q (T_B)}{d
T_B}\frac{\nu_{q}}{p+\nu_q}\Bigr]}.
\end{equation}
We choose the value $ \sigma $ so that all singularities of the integrand are to the left of the
path of integration.

Substituting Eq.~(\ref {3}) in Eq.~(\ref {18})  and integrating over the angles of the vector $
{\bf q} $, we obtain

\begin{eqnarray}\label{19}
&\Delta & T_e (t)= \frac{1}{2\pi i} \int_{\sigma-i\infty}^{\sigma+i\infty}dp\,\, e^{pt} \Delta T_e
(0) p^{-1}
 \Bigl\{1 \nonumber \\
 &+&\frac{1}{\zeta_1}\int_{0}^{\infty}{{ x^{5+r}}\over{{\rm sinh}^{2}(x/2)[x^{1+r}+p/\nu(T_B)]}}
dx\Bigr\}^{-1},
\end{eqnarray}
where the dimensionless variable $ x = \hbar \omega_{q} / T_B $, and the characteristic collision
frequency of thermal phonons with electrons $ \nu (T_B) = {{m^2 \mu_{r}^{2}} \over { 2 \pi \hbar^4
\rho_f s}} T_{B}^{1 + r} $. The constant $ \zeta_1 = 16 \pi^4 c_e / (15 c_p) $ is determined by
the ratio of the electron and phonon heat capacities at the temperature of the thermostat. For the
phonon heat capacity in the model with one (acoustic) mode of oscillations, we have the expression
$ c_p = (2 \pi^2 T_{B}^{3}) / (15 \hbar^3 s^3) $. It is not difficult to see that the case $ r =
-1 $ (for which $ \tau_ {e}^{-1} \propto T_{e}^{2} $) is special. Indeed, integrations over $ x $
and $ p $ are separated for $ r = -1 $ only, and this, as will be seen later, leads to an
exponential dependence of $ T_ {e} $ on time.  The case $ r = -1 $ will be considered separately,
but for now we shall assume $ r \neq -1 $.

To calculate the integral over $ p $ in  Eq.~(\ref {19}), we need to continue analytically the
integrand to the region located to the left of the integration contour, i.e. into the domain $
{\rm Im} (p) <\sigma $. The function of the complex variable $ p $, given by an integral over $ x
$, has singularities for negative values of $ p $. An analytic continuation of such a function is
performed by cutting a complex $ p $-plane along the negative part of the real axis. In this case,
the integral over $ p $ reduces to the contribution of the pole $ p = 0 $ and to integrals along
the edges of the cut. Making the substitutions $ x^{1 + r} = y $, $ p = -p_1 \nu (T_B) $ and using
the well-known formula

\begin{equation}\label{20}
\lim_{\delta\to{+0}} \int \frac{f(y)dy}{y-y_0 \pm i\delta} = V.p.\int \frac{f(y)dy}{y-y_0 } \mp i
\pi f(y_0),
\end{equation}
we obtain

\begin{widetext}
\begin{equation}\label{21}
\Delta T_e (t)= \frac{c_e \Delta T_e (0)}{c_e+c_p} + \int_{0}^{\infty} \frac{\zeta_1 (1+r)
p_{1}^{(4-r)/(1+r)} e^{-p_1 t\nu(T_B)}\Delta T_e (0)dp_1} { \sinh^{2}(\frac{p_{1}^{1/(1+r)}}{2})
\Bigl\{ \Bigl[\zeta_1 (1+r)+
V.p.\int_{0}^{\infty}\frac{y^{5/(1+r)}dy}{(y-p_1)\sinh^{2}(y^{1/(1+r)}/2)}\Bigr]^2+\frac{\pi^2
p_{1}^{10/(1+r)}}{\sinh^{4}(p_{1}^{1/(1+r)}/2)}\Bigr\}}.
\end{equation}
\end{widetext}
From  Eq.~(\ref {21}), it is not difficult to find the behavior of $ \Delta T_e (t) $ for large
times, when $ t \nu (T_B) \gg 1 $. In this limit, the main contribution to the integral over $ p_1
$ is given by $ p_1 \ll 1 $, that allows us to neglect $ p_1 $ in the integral over $ y $, and
neglect the  last term in curly brackets. Replacing the hyperbolic sine by its argument, we obtain
the expression

\begin{eqnarray}\label{22}
&\Delta& T_e (t) \approx \frac{c_e \Delta T_e (0)}{c_e+c_p}
+\frac{5}{4\pi^4}\Gamma\Bigl(\frac{4+r}{1+r}\Bigr)\frac{c_e c_p\Delta T_e (0)}{(c_e+c_p)^2}
\nonumber\\
 &\times& [t\nu(T_B)]^{-3/(1+r)},
\end{eqnarray}
where $ \Gamma $ is the gamma function. It follows from Eq.~(\ref {22}) that for large times the
relaxation of the electron temperature has a power-law character. The term $ c_e \Delta T (0) /
(c_e + c_p) $ determines the electron temperature, which is established in a thermally insulated
film at $ t \to \infty $, when electrons and phonons come to thermodynamic equilibrium.

For small times, the asymptotic behavior of $ \Delta T_e (t) $ is convenient to find  from
Eq.~(\ref {18}) by expanding the integrand in powers of $ p^{-1} $. If we retain the first two
terms of the expansion, we obtain

\begin{equation}\label{23}
\Delta T_e (t)\approx \Delta T_e (0)(1-t/\tau_e),
\end{equation}
where the average frequency of electron-phonon collisions

\begin{equation}\label{24}
\tau_{e}^{-1} = \frac{1}{c_e}\int{{d^3 q}\over{(2\pi)^3}}\hbar\omega_{q} \nu_q \frac{d n_q
(T_B)}{d T_B}.
\end{equation}

In our opinion, the slow power-law relaxation of the electron temperature in thermally insulated
films at large times is a very unexpected result of Sec.~\ref {s3}, since this behavior of $
\Delta T_e (t) $ is qualitatively different from the exponential relaxation predicted by 2TM (see
the Appendix).

 \section{Heat removal to substrate, and three-temperature model}\label{s4}

The removal of heat from the  film to the substrate substantially complicates the picture of
cooling of the heated electrons. With a relatively weak heating and $ \alpha \neq 0 $, the
dynamics of the electron temperature is described by a linear equation (\ref {12}), which can be
solved by the operational method, just like Eq.~(\ref {13}). For the Laplace transform of the
deviation of the electron temperature from $ T_B $, we obtain

\begin{eqnarray}\label{25}
&\Delta& \tilde{T}_e (p) =  c_e \Delta T_e (0)\Bigl\{ c_e p + 2\int_{q_z >0} {{d^3 q}\over
{(2\pi)^3}} \hbar\omega_{q}\,\nu_q {{dn_q(T_B)}\over {dT_B}}  \nonumber \\
&\times& {\frac{1}{p+\nu_q}} \Bigl [p + {\frac{\alpha
s_z\nu_q}{2d(p+\nu_q)}{\frac{1-\chi(p)}{1-\beta\chi(p)}}} \Bigr ] \Bigr\}^{-1},
\end{eqnarray}
with $ \chi (p) = \exp [-2d (p + \nu_q) / s_z] $. We notice immediately that in the case $ \alpha
\neq 0 $ Eq.~(\ref {25}) has two new features compared to $ \Delta \tilde {T}_e (p) $, obtained in
Sec.~\ref {s3}. First, the pole $ p = 0 $ disappears, so that  $ \Delta T_e (t) \to 0$ as $ t \to
\infty $. Second, Eq.~(\ref {25}) contains a new parameter, namely, the film thickness $ d $.
Further we shall see that the nature of the thermal relaxation of the electronic subsystem depends
essentially on the value of $d $. Since Eq.~(\ref {25}) describes the relaxation of the electron
temperature for systems with arbitrary $ r $ and arbitrary parameter values, it turns out to be
rather complicated for analysis. Such an analysis is greatly simplified for $ r = -1 $, when the
frequency of phonon-electron collisions does not depend on the wave vector of the phonon. It is
the  case that will be considered in the next Subsection. Recall that $ r = -1 $ corresponds to
the dependence $\tau_{e}^{-l} \propto T_{e}^{2} $, which is often observed in the experiments
\cite{LinJPhysCondMatt2002, KarasikIEEEApplSupercond2007, WeiNatureNanotechnol2008,
GershensonJETP1990, BergmannPRB1990}.

\subsection{The frequency of phonon-electron collisions does not depend on the wave vector of the phonon}\label{ss4.1}

It follows from Eq.~(\ref {3}) that the frequency of phonon-electron collisions $ \nu_q $ loses
its dependence on the phonon wave vector $ {\bf q} $ for $ r = -1 $. This circumstance
substantially facilitates the integration over modulus $ {\bf q} $. To simplify  the integration
over the angles of the vector $ {\bf q} $, we assume that the probability  $ \alpha (\theta) $ is
equal to the constant $ \alpha_0 $ for $ \theta <\theta_{cr} $ and is equal to zero for $ \theta_
{cr} <\theta <\pi / 2 $. Recall that the angle $\theta_{cr} $ is the angle of total internal
reflection, which is determined by the sound velocities in the film and in the substrate according
to the equation $ \sin (\theta_{cr}) = (s / s ^{\prime} ) $. The obtained step approximation has
all the characteristic features of the real dependence $ \alpha (\theta) $ (see Fig.~\ref {f1}) if
we take $ \alpha_0 \approx \alpha (0) $. Now we take into account that for sufficiently different
velocities $ s $ and $ s ^{\prime} $ the angle $ \theta_{cr} $ is not too close to $ \pi / 2 $,
and hence the projection of the phonon velocity $ s_z $ of the order $ s $. Replacing $ s_z $ by $
s $  and using the stepwise approximation $ \alpha (\theta) $, we obtain for $ r = -1 $

\begin{widetext}
\begin{equation}\label{26}
\Delta \tilde{T}_e (p) =  c_e \Delta T_e (0)\Bigl\{ c_e p + c_p {{p\nu_0}\over {(p+\nu_0})}+ c_p
{{\nu_0^2}\over {(p+\nu_0)^2}} \,{{(1-\eta_0)\alpha_0 s}\over
{2d}}\,{\frac{[1-\exp(-2d(p+\nu_0)/s))]}{[1-\beta_0\exp(-2d(p+\nu_0)/s)]}}\Bigr\}^{-1}.\,\,\,\,\,\,
\end{equation}
\end{widetext}
Here $ \nu_0 $ is the frequency of phonon-electron collisions that does not depend on the wave
vector of the phonon: $ \nu_0 = m^2 \mu_{- 1}^{2} / (2 \pi \hbar^4 \rho_f s) $.

We denote the fraction of ("trapped") phonons with angles $ \theta_{cr} <\theta <\pi / 2 $ via $
\eta_0 $. A simple calculation of this fraction  yields $ \eta_0 =  \cos (\theta_{cr}) $.
Accordingly, the fraction of phonons that can leave the film is $ 1- \eta_0 $. It is this value
that appears in Eq.~(\ref {26}).

In order to  distinguish limiting cases, we introduce the dimensionless quantities $ \bar p = p /
\nu_0 $, $ \bar d = 2d \nu_0 / s $, and $ \zeta = c_e / c_p $. The Laplace transform of the
electron temperature takes the form

\begin{equation}\label{27}
\Delta \tilde{T}_e (\bar p)=\zeta\Delta T_e (0)/[\nu_0 F(\bar p)],
\end{equation}
with

\begin{equation}\label{28}
F(\bar p) =  \zeta\bar p + {{\bar p}\over {(\bar p+1})}+ {{(1-\eta_0)\alpha_0}\over {(\bar p+1)^2
\bar d}}{\frac{\{1-\exp[-\bar d(\bar p+1)]\}}{\{1-\beta_0\exp[-\bar d(\bar p+1)]\}}}.
\end{equation}

\begin{figure}[t]
\includegraphics[width=9.0 cm]{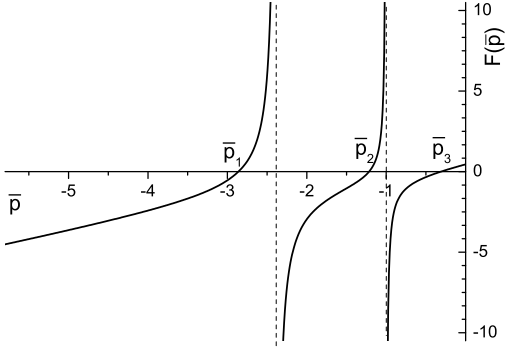}
\caption{\label{f2}The graph of the function $ F (\bar p) $ for $ \zeta = 1 $, $ \eta_0 = 0.2 $,
  $ \bar d = 0.5 $,
$ \alpha_0 = $ 0.5. Three zeros of $ F (\bar p) $, which are denoted by $ \bar {p}_1 $, $ \bar
{p}_2 $, $ \bar {p}_3 $, give the three poles $ \Delta \tilde {T}_e (p) $. }
\end{figure}

\begin{figure}[t]
\includegraphics[width=9.0 cm]{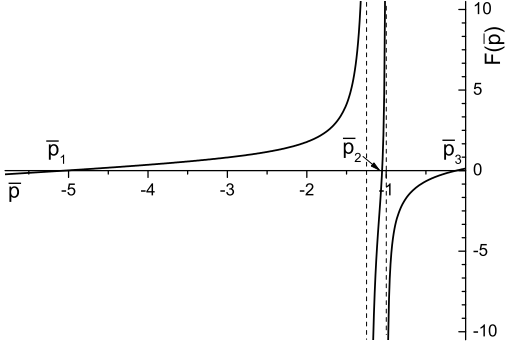}
\caption{\label{f3}The graph of the function $ F (\bar p) $ for $ \zeta = 0.25 $, $ \eta_0 = 0.2
$, $ \bar d = 3 $, $ \alpha_0 = $ 0.5. Compared to Fig.~\ref {f2}, the pole $ \bar{p}_1 $ moved to
the left in the region of large values $ \vert \bar{p} \vert $, and the pole $ \bar{p}_3 $ shifted
to the right close to the point $ \bar{p} = 0 $.
 }
\end{figure}

The graph of $ F (\bar p) $ for fairly general (not large or small) values of the parameters ($
\zeta = 1 $, $ \eta_0 = 0.2 $, $ \bar d = 0.5 $, $ \alpha_0 = 0.5 $) is shown in Fig.~\ref {f2}.
We see that the function $ F (\bar p) $ has three zeros. These zeros correspond to the poles of
the complex function $ \Delta \tilde {T} (p) $, which means that the dependence $ \Delta {T} (t) $
will be a linear combination of three exponentials with exponents $ (\bar {p}_i t \nu_0) $, where
$ i = 1,2,3 $. To find out the meaning of the poles, consider how $ F (\bar p) $ changes with
decreasing $ \zeta $ and with increasing $ \bar d $. Let us take $ \zeta = 0.25 $ and $ \bar d = 3
$, and leave the rest of the parameters unchanged. The dependence $ F (\bar p) $ for new parameter
values is presented in Fig.~\ref {f3}. It is clearly seen (and this will be shown analytically)
that the decrease of $ \zeta $ strongly affects the pole $ \bar {p}_1 $, shifting it to the left
in the region of large values $ \vert \bar {p} \vert $, while an increase of $ \bar d $ shifts the
pole $ \bar {p}_3 $ to the right to the point $ \bar{p} = 0 $.

For small $ \zeta $, the pole $ \bar p_1 $ must be searched in the domain $ \vert \bar {p} \vert
\gg 1 $. In this region, we have $ F (\bar p) \approx \zeta \bar {p} + 1-1 / \bar {p} $. The root
of this expression is $ \bar p_1 \approx -1 / \zeta-1 $, whence it is clear that small $ \zeta $
corresponds to large $ \vert \bar {p}_1 \vert $. The pole $ \bar p_1 $ describes the initial stage
of relaxation of the electron temperature, when hot electrons  emit nonequilibrium phonons. This
stage is characterized by the exponential $ e^{\bar {p}_1 \nu_0 t} = e ^{- (\nu_0 + 1 / \tau_e) t}
$.  Here we used the equality $ \nu_0 c_p / c_e = 1 / \tau_e $, from which it follows that for $
\zeta \ll 1 $ the relaxation is determined by the time $ \tau_e $, since $ \tau_e \ll 1 / \nu_0 $.

The pole $ \bar p_3 $ is easy to find if we take into account that for large film thicknesses $
\bar d \gg 1 $ the value $ \vert \bar {p}_3 \vert \ll~1 $. In the domain of small $ \vert \bar {p}
\vert $, the function $ F (\bar p) \approx (\zeta +1) \bar {p} + (1- \eta_0) \alpha_0 / \bar {d}
$, from which we have $ \bar {p}_3 = - (1- \eta_0) \alpha_0 / [(\zeta +1) \bar {d}] $. The pole $
\bar p_3 $ describes the late stages of relaxation of the electron temperature. The corresponding
time dependence has the form $ \Delta {T}_e (t) \propto e^{-t / \tau_{es}^{\prime}} $, where $
\tau_{es}^{\prime} = (1- \eta_0) \alpha_0 s / [(\zeta +1) 2 d] $. We notice that the time $
\tau_{es}^{\prime} $ is proportional to the characteristic time of phonon escape from the film $
\tau_{es} = \alpha_0 s / (2d) $. We also notice that unlike the time $ \tau_{es} $, which
characterizes a single phonon, the relaxation time of the electron temperature $
\tau_{es}^{\prime} $ contains the factor $ (1- \eta_0) $, which takes into account only that part
of the phonons, which in principle can leave the film, and the factor $ (\zeta +1) $, which takes
into account the thermal inertia of the electrons.

To find the pole $ \bar p_2 $, we take into account that for $ \bar d \gg 1 $ the inequality $
(\bar p_2 + 1)  \ll 1 $ holds. Expanding $ F (\bar p) $ with a small value $ (\bar p_2 + 1) $, we
obtain

\begin{equation}\label{29}
\bar p_2 = -1- 2\alpha_0\eta_0 /[(1- \eta_0)(1+\beta_0)\bar d].
\end{equation}

The time dependence of the electron temperature is given by the inverse  Laplace transformation.
Since for $ r = -1 $ the singular points (simple poles) of the integrand are the roots of $ F
(\bar p) $, the result of the inverse transformation can be written as the sum

\begin{equation}\label{30}
\Delta T_e (t) = \sum_{i=1}^{3}\frac{\zeta \Delta T_e (0)}{F^\prime (\bar p_i)}e^{\bar p_i \nu_0
t},
\end{equation}
with $ F^\prime (\bar p_i) = \frac {d F (\bar p)} {d \bar p} \mid_{\bar p = \bar p_i} $. It is
seen that the larger the derivative $ F^\prime (\bar p_i) $, the smaller the contribution to $
\Delta T_e (t) $ is given by the corresponding pole $\bar p_i$.

\subsection{Three-temperature model}\label{ss4.2}

The microscopic analysis of the relaxation of the electron temperature, carried out in
Sec.~\ref{ss4.1}, showed that for $ r = -1 $ the relaxation process is described by a linear
combination of three exponentials. A different result, namely, the dependence $ \Delta {T_e} (t) $
is determined by two exponentials, was obtained in  Appendix, where the electron temperature
relaxation is analyzed on the basis of 2TM. The latter is not surprising since in 2TM the energy
relaxation is described by only two linear differential equations for the electron and phonon
temperatures. The reason for the qualitative difference between the two relaxation scenarios is
quite simple: 2TM does not take into account the fact that the phonons are divided into two groups
by their ability to pass through the interface between the film and the substrate. In fact, if
phonons with angles of incidence $ \theta <\theta_{cr} $ can pass into the substrate with a
probability $ \alpha_0 $, then the phonons with $ \theta_{cr} <\theta <\pi / 2 $ are completely
reflected from the interface. In this Subsection we consider a three-temperature model (3TM),
which assigns the temperature $ T_e $ to the electrons and assigns the temperatures $ T_ {p1} $
and $ T_ {p2} $ to the phonon groups noticed above. 3TM gives the three-exponential behavior of $
\Delta {T_e} (t) $, which improves the description of the relaxation of the electron temperature
and, in a number of cases, yields results that are practically not different from the results of
the microscopic model (MM) presented in Sec.~\ref{ss4.1}.

Thus, within the framework of 3TM, the process of relaxation of the electron-phonon system in a
metallic film is described by the following set of equations:

\begin{equation}\label{b1}
c_e(T_e)\frac{dT_e}{dt}= -\Sigma_5 (1-\eta_0)(T_e^5-T_{p1}^5)-\Sigma_5 \eta_0(T_e^5-T_{p2}^5),
\end{equation}

\begin{equation}\label{b2}
c_p(T_{p1})\frac{dT_{p1}}{dt}=  \Sigma_5 (T_e^5-T_{p1}^5)-K(T_{p1}^4-T_B^4),
\end{equation}

\begin{equation}\label{b3}
c_p(T_{p2})\frac{dT_{p2}}{dt}=  \Sigma_5 (T_e^5-T_{p2}^5),
\end{equation}
with $ \eta_0 = \cos (\theta_{cr}) $. The constants $ \Sigma_5 $ and $ K $ describe the heat
transfer between electrons and phonons in the film and between the phonon subsystems of the film
and the substrate, respectively. (For simplicity, we consider the case of pure films.) If the
initial heating of the electrons is small, the set of equations can be linearized in $ \Delta T_e
$, $ \Delta T_{p1} $ and $ \Delta T_{p2} $, which are small deviations of the temperatures $ T_e
$, $ T_{p1} $, and $ T_{p2} $ from the bath temperature  $ T_B $. Introducing the energy
relaxation time of electrons $ \tau_e = c_e / (5 T_B^4 \Sigma_5) $ and the time of phonon escape
from the film $ \tau_{es} = c_p / (4 K T_B^3) $, we obtain a set of linear equations

\begin{equation}\label{b4}
\frac{d\Delta T_e}{dt}= -\frac{1}{\tau_e} (1-\eta_0)(\Delta T_e - \Delta T_{p1}) -\frac{1}{\tau_e}
\eta_0(\Delta T_e - \Delta T_{p2}),
\end{equation}

\begin{equation}\label{b5}
\frac{d\Delta T_{p1}}{dt}=  \frac{c_e}{c_p\tau_e} (\Delta T_e-\Delta
T_{p1})-\frac{1}{\tau_{es}}\Delta T_{p1},
\end{equation}

\begin{equation}\label{b6}
\frac{d\Delta T_{p2}}{dt}=  \frac{c_e}{c_p\tau_e} (\Delta T_e-\Delta T_{p2})
\end{equation}
with the initial conditions $ \Delta T_e (t = 0) = \Delta T_e (0) $ and $ \Delta T_{p1} (t = 0) =
\Delta T_{p2} (t =~0) = 0 $. In order to be able to compare the predictions of 3TM  with the
results of the microscopic model and 2TM, we solve the set of equations (\ref{b4}), (\ref{b5}),
(\ref{b6}) by the operational method. Introducing $ \nu_e = 1 / \tau_e $, $ \nu_p = c_e \nu_e /c_p
$ and $ \nu_{es} = 1 / \tau_ {es} $, for Laplace transforms $ \Delta \tilde T_e (p ) $, $ \Delta
\tilde T_{p1} (p) $ and $ \Delta \tilde T_{p2} (p) $, we obtain

\begin{equation}\label{b7}
\Delta \tilde T_e(p)=  (p+\nu_p)(p+\nu_p+\nu_{es})\Delta T_e(0)/D,
\end{equation}

\begin{equation}\label{b8}
\Delta \tilde T_{p1}(p)=  (p+\nu_p) \nu_p \Delta T_e(0)/D,
\end{equation}

\begin{equation}\label{b9}
\Delta \tilde T_{p2}(p)=  \nu_p(p+\nu_p+\nu_{es})\Delta T_e(0)/D,
\end{equation}
with the following determinant of the set of equations:

\begin{eqnarray}\label{b10}
&D& = p^3 + (\nu_{e}+2\nu_p+\nu_{es})p^2 + (\nu_p+\nu_{es})(\nu_p+\nu_{e})p\nonumber \\
 &+&
(1-\eta_0)\nu_{e}\nu_p\nu_{es}.
\end{eqnarray}

In the low-temperature region, where the phonons are not thermalized the microscopic theory
describes the relaxation of the electron-phonon system in terms of the electron temperature and
the phonon distribution function. We notice that the phenomenological models (3TM and 2TM) also
give their own expressions for the electron temperature, in which phenomenological parameters
appear. Comparing these expressions with the results of the microscopic theory [see
Eqs.~(\ref{27}), (\ref{28})], we can relate microscopic and phenomenological parameters, and then
compare the predictions of MM with 3TM and 2TM.

In order to express the phenomenological frequencies $ \nu_e $ and $ \nu_p $ via microscopic
parameters, it is convenient to take the limiting case of a thermally insulated film. For $
\nu_{es} $, it is convenient to use the case of a weak thermal contact between the film and the
substrate. Comparing Eqs.~(\ref{27}), (\ref{28}) with Eqs.~(\ref{b7}), (\ref{b10}), it immediately
follows that $ \nu_p = \nu_0 $, and $ \nu_e = c_p \nu_0 / c_e $. For a bad heat sink (when $
\nu_{es} \ll \nu_e, \nu_p $), the determinant $ D $ has the root $ p_3 = - (1- \eta_0) \nu_{es}
\nu_{e} / (\nu_{e} + \nu_{p}) $. Comparing it with the root $ \bar p_3 $, found in
Sec.~\ref{ss4.1}, we have $ \nu_{es} = \alpha_0 s / (2d) $.

Now we can compare the dynamics of relaxation of the electron temperature, obtained on the basis
of the microscopic theory, with the results of 3TM and 2TM. To do this, we rewrite the Laplace
transforms of the electron temperature, given by Eqs.~(\ref{b7}) and (\ref{b10}), in the
dimensionless quantities introduced in Sec.~\ref{ss4.1}. We have for 3TM


\begin{widetext}
\begin{equation}\label{b11}
\Delta \tilde T_e(\bar p)= \frac{(\bar p+1)(\bar p+1+\alpha_0/\bar d)\zeta \Delta
T_e(0)}{\nu_0[\zeta\bar{p}^3+ (1+2\zeta+\alpha_0\zeta/\bar d)\bar{p}^2+(1+\zeta)(1+\alpha_0/\bar
d)\bar{p}+(1-\eta_0)\alpha_0/\bar d]}.
\end{equation}
\end{widetext}
A similar transformation of Eq.~(\ref{a5}) yields for 2TM

\begin{equation}\label{b12}
\Delta \tilde T_e(\bar p)= \frac{(\bar p+1+\alpha_0/\bar d)\zeta \Delta
T_e(0)}{\nu_0[\zeta\bar{p}^2+ (1+\zeta+\alpha_0\zeta/\bar d)\bar{p}+\alpha_0/\bar d]}.
\end{equation}

\begin{figure}[t]
\includegraphics[width=9.0 cm]{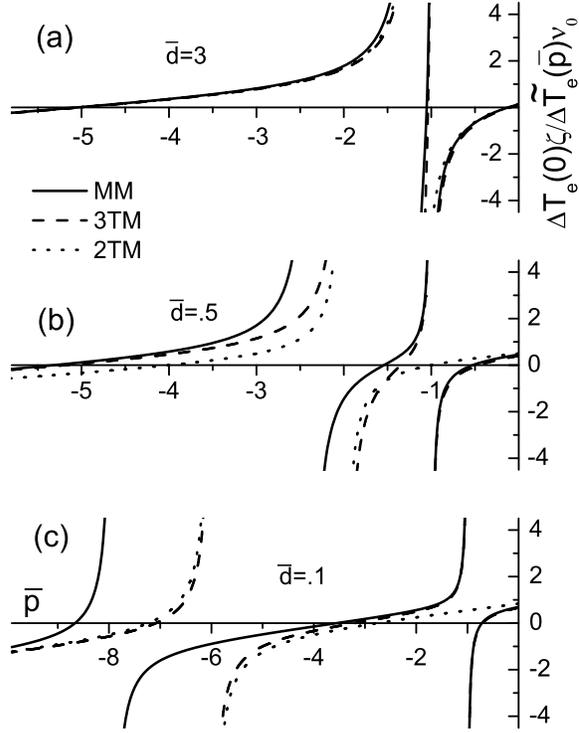}
\caption{\label{f4}Dependence of $ \Delta T (0) \zeta / [\Delta \tilde T_e (\bar p) \nu_0] $ on $
\bar p $ for films of different thicknesses: (a) $ \bar d = 3 $; (b) $ \bar d = 0.5 $; (c) $ \bar
d = 0.1 $. The solid lines represent the function $ F (\bar p) $, obtained on the basis of MM
Eq.~(\ref{28}), dashed and dotted lines give the results of 3TM Eq.~(\ref{b11}) and 2TM
Eq.~(\ref{b12}), respectively . The parameters in Eqs.~(\ref{28}), (\ref{b11}), and (\ref{b12})
have the following values: $ \alpha_0 = 0.5 $, $ \eta_0 = 0.2 $, $ \zeta = 0.25 $. Notice that the
solid and dashed curves differ slightly for $\bar p > -1$.
 }
\end{figure}

Comparison of the predictions of the three models is much simpler and more obvious if we consider
not the time dependence of the electron temperature, but the dependence of $ \Delta T (0) \zeta /
[\Delta \tilde T_e (\bar p) \nu_0] $ on the dimensionless variable $ \bar p $. Fig.~\ref {f4}
shows the dependence of $ \Delta T (0) \zeta / [\Delta \tilde T_e (\bar p) \nu_0] $ on $ \bar p$
for films of different thicknesses. The results of MM  are given by a solid line, and the results
of 3TM and 2TM are given by dashed and dotted  lines, respectively. It is seen from Fig.~\ref
{f4}(a) that for effectively thick films ($ d_{ef} \gg 1 $; $ d_{ef} \equiv \bar d / \alpha_0 $),
the results of MM and 3TM are practically coincide, whereas 2TM does not reproduce the root $\bar
 p_2 $. As for films with a thickness of $ d_{ef} = 1 $ [see Fig.~\ref {f4}(b)], such films are
fairly well described by 3TM, but very badly by 2TM.  Only for effectively thin films with $
d_{ef} \ll~1 $, 3TM does not satisfactorily describe the early stages of the relaxation of the
electron temperature. However, 3TM describes these stages no worse than 2TM. Thus, we can conclude
that, in comparison with the well-known 2TM, the phenomenological 3TM, which takes into account
the locking (in the film) of phonons with angles of incidence $ \theta >\theta_ {cr} $, improves
the description of the electron temperature relaxation.

\subsection{Kinetics of electron cooling in pure films} \label{ss4.3}

In Sec.~\ref{ss4.2}, we examined the case $ r = -1 $ when the frequency of phonon-electron
collisions $ \nu_q $ does not depend on the wave vector of the phonon. Now we consider the more
general situation when $ \nu_q $ depends on $ q $. Let us take a case of pure, i.e. $r=0$. In this
case, the dependence of $ \nu_q $ on $ q $ is linear, and this simplifies calculations and gives a
more clear picture of the energy relaxation of electrons without cluttering it with nonessential
details. As in Sec.~\ref{ss4.1}, we will consider the dependence $ \alpha (\theta) $ as stepwise,
so that $ \alpha = \alpha_0 $ for $ \theta <\theta_{cr} $ and $ \alpha = 0 $ for $ \theta>
\theta_{cr} $. In addition, we set $ s_z = s $. Under these assumptions, the Laplace transform of
the deviation of the electron temperature has the form

\begin{widetext}
\begin{equation}\label{43}
\Delta \tilde{T}_e (p_1) =   \frac{\Delta T_e (0)}{\nu} \Bigl\{  p_1 + \frac{1}{\zeta_1}
\int_{0}^{\infty}{{x^4 dx}\over {\sinh^2(x/2)}}{{p_1 x}\over {(p_1+x)}} +
\frac{(1-\eta_0)\alpha_0}{\zeta_1 d_1} \int_{0}^{\infty}{{x^4 dx}\over {\sinh^2(x/2)}}{{ x^2}\over
{(p_1+x)^2}} {\frac{[1-\exp(-d_1(p_1+x))]}{[1-\beta_0\exp(-d_1(p_1+x))]}}\Bigr\}^{-1},
\end{equation}
\end{widetext}
with dimensionless quantities $p_1=p/\nu(T_B)$, $x=\hbar\omega_q/T_B$, $d_1=2d\nu(T_B)/s$ ш
$\zeta_1=16\pi^4 c_e/(15 c_p)$.

The inverse Laplace transformation of Eq.~(\ref{43}) leads to an integral along the edges of the
cut in the complex $ p_1 $-plane passing along the negative part of the real axis. This integral
can be written as

\begin{equation}\label{44}
\Delta T_e (t)= - \frac{1}{\pi}\int_{-\infty}^{0}dp_1\frac{e^{p_1 t\nu}\Delta T_e
(0)\Phi''(p_1)}{[\Phi'(p_1)]^2 + [\Phi''(p_1)]^2},
\end{equation}
where $ \Phi'(p_1) $ and $ \Phi''(p_1) $ are the real and imaginary parts of the expression
enclosed in curly brackets in Eq.~(\ref{43}). The calculation of these quantities gives

\begin{widetext}
\begin{eqnarray}\label{45}
\Phi'(p_1) = p_1+     \frac{1}{\zeta_1}V.p. \int_{0}^{\infty}{{x^4 dx}\over {\sinh^2(x/2)}}{{p_1
x}\over {(p_1+x)}}
 -\frac{(1-\eta_0)\alpha_0}{\zeta_1 \ln (\beta_0)} V.p. \int_{0}^{\infty}{{x^6
dx}\over {\sinh^2(x/2)}}{{ R(x,p_1)}\over {(p_1+x)}}\nonumber \\
+
\frac{(1-\eta_0)\alpha_0}{\zeta_1 \ln (\beta_0)} V.p. \int_{0}^{\infty}{{x^6 dx}\over
{\sinh^2(x/2)}}{{ R(x,p_1)}\over {(p_1+x-\frac{1}{d_1} \ln (\beta_0))}},
\end{eqnarray}

\begin{equation}\label{46}
\Phi''(p_1) =- \frac{\pi \eta_0}{\zeta_1}\,{{p_{1}^{6}}\over {\sinh^2 (p_1/2)}} -
 \frac{\pi(1-\eta_0)\alpha_{0}^{2}}{\zeta_1 \beta_0\ln ^2 (\beta_0)}\,\,
 {{(\frac{1}{d_1} \ln (\beta_0)-p_1)^6 \,\Theta (\frac{1}{d_1} \ln (\beta_0))}
 \over {\sinh^2[\frac{1}{2}(\frac{1}{d_1} \ln (\beta_0)-p_1)]}}.
\end{equation}
\end{widetext}
Here $ \Theta (x) $ is the Heaviside step function, and the function $ R (x, p_1) $, which does
not have singularities, is defined by the following expression:

\begin{equation}\label{47}
 R(x,p_1) = \frac{p_1+x- \frac{1}{d_1} \ln (\beta_0)}{p_1+x}\,\frac{[1-\exp(-d_1(p_1+x))]}
 {[1-\beta_0\exp(-d_1(p_1+x))]}.
\end{equation}

To identify the processes that determine the physics of energy relaxation in a metal film lying on
a dielectric substrate, we consider the limiting cases of thick and thin films, as well as early
and late stages of relaxation. First, let us consider the dynamics of the electron temperature in
a thick film ($ d_1 \gg 1 $) in the late stages ($ t \nu \gg 1 $). For $ t \nu \gg 1 $, the main
contribution to the integral (\ref{44}) is made by $ | p_1 | \ll 1 $, while the characteristic $ x
$ in Eq.~(\ref{45}) is of the order of unity. As a consequence, Eq.~(\ref{45}) reduces to the form

\begin{equation}\label{48}
\Phi'(p_1) = (1+\frac{c_p}{c_e})\,(p_1+\frac{1}{\tau'_{es} \nu}),
\end{equation}
where

\begin{equation}\label{49}
\tau'_{es} =  \frac{(1-\eta_0)\alpha_{0} s}{2d}\,\frac{c_p}{c_p+c_e}.
\end{equation}

Under the assumption that the thermal contact of the film with the substrate is not very good, so
that $ \tau'_{es} \nu \gg 1 $, the dependence of the electron temperature on time has an
exponential character,

\begin{equation}\label{50}
\Delta T_e (t)=  \frac{c_e\Delta T_e (0)}{c_p+c_e} e^{-t/\tau'_{es}}.
\end{equation}

It should be noticed that Eq.~(\ref{50}) is valid if the time $ t$ is of the order of $ \tau'_{es}
$. If $ t \gg \tau'_{es} $, we obtain the power-law dependence

\begin{equation}\label{51}
\Delta T_e (t)=  \frac{4\Gamma (5)\eta_0 \Delta T_e (0)}{\zeta_1(1+\frac{c_p}{c_e})^2}\,
\frac{1}{(t\nu)^3 (t/\tau'_{es})^2}.
\end{equation}

The reason for changing the asymptotic behavior of $\Delta T_e (t)$ is fairly simple. If
Eq.~(\ref{50}) represents a slow cooling of a system of electrons and phonons that have come into
equilibrium with each other, Eq.~(\ref{51}) describes the relaxation of phonons, which initially
had angles of incidence $ \theta > \theta_{cr} $, i.e. they were originally  "locked" in  the
film. Such phonons can leave the film  in the following way: first they are again absorbed by
electrons, and then emitted into the angular range $0 < \theta < \theta_{cr} $. Additional
evidence in favor of the fact that the asymptotic behavior (\ref{51}) is related to the group of
"locked" phonons, is its proportionality $ \eta_0 $. If $ \eta_0 = 0 $, that is, if all phonons
have the ability to leave the film without reabsorbing by electrons, such  asymptotic behavior do
not arise.

In the late stages, the exponential relaxation of the electron temperature takes place in thin
films, in which $ d_1 \ll \alpha_0 / \beta_0 $. If $ t $ is greater, but of the order $ \tau_e $,
and moreover $ \eta_0 \ll 1 $, then

\begin{equation}\label{52}
\Delta T_e (t)= \Delta T_e (0) e^{-t/\tau_{e}},
\end{equation}
where the average frequency of electron-phonon collisions $\tau_{e}^{-1}$ is determined by
Eq.~(\ref{24}).

The last stage (associated with the escape from the film of phonons with angles of incidence $
\theta > \theta_{cr} $ and which takes place when $ t \gg \tau_{e} $) is characterized by the
following power-law dependence:

\begin{equation}\label{53}
\Delta T_e (t)=  \frac{4\Gamma (5)\eta_0 \Delta T_e (0)}{\zeta_1}\, \frac{1}{(t\nu)^3
(t/\tau_{e})^2}.
\end{equation}

In the early stages of relaxation, the heat transfer from the film is unimportant, and therefore
for both thick and thin films the dependence of the electron temperature on time is given by
Eq.~(\ref{23}).

We notice that Eq.~(\ref{50}) gives a condition under which the thick film can be considered as
thermally
 insulated one. Namely, we can neglect the heat removal to the substrate if the energy relaxation
in the electron-phonon system of the film is considered at times substantially shorter than the
time $ \tau'_{es} $. Thus, the results of the Section~\ref{s3} are valid in this time domain under
the condition  $ \tau'_{es} \gg \tau_{e} $ and $ \nu (T_B) \tau'_{es } \gg 1 $.

\section{Discussion  and conclusions} \label{s5}

At present, the two-temperature model, in which electrons and phonons are characterized by their
temperatures, is mainly used to describe the thermal relaxation between heated electrons and a
cold lattice. The microscopic theory of cooling of electrons presented in Sec.~\ref{s2} overcomes
the limitation of 2TM, which does not take into account a significant group of nonequilibrium
phonons that cannot "fly" out of the metal film into the substrate, because their incidence angles
$ \theta $ are larger the critical angle $ \theta_ {cr} $.  We notice, however, that the
microscopic theory (where electrons are characterized by the temperature $ T_e $ and phonons are
described by a distribution function) leads to a rather complex integro-differential equation
 for $ T_e (t) $, which makes it difficult to use. A simpler phenomenological allowance
for phonons with $ \theta > \theta_{cr} $ can be based on the introduction of an intrinsic
temperature for them, and therefore the phenomenological model that more adequately describes the
energy relaxation in a thin-film system turns out to be a three-temperature one
(Sec.~\ref{ss4.2}). The three-temperature model improves the description of the energy relaxation
(in comparison with 2TM).

At the same time, 3TM has its own field of application as it can be seen from the results obtained
in the Sec.~\ref{s3} and Sec.~\ref{ss4.3}. 3TM most accurately describes the case of $ r = -1 $,
when the frequency of phonon-electron collisions does not depend on the phonon wave vector
(Sec.~\ref{ss4.1}). Recall that the case of $ r = -1 $ corresponds to dirty metal films where
electrons are mainly scattered on heavy impurity atoms or grain boundaries, which are weakly
entrained by lattice oscillations. Such scattering leads to the dependence $ \tau_{e}^{-1} \propto
T_{e}^{2} $, which is often observed in  experiments \cite{LinJPhysCondMatt2002,
KarasikIEEEApplSupercond2007, WeiNatureNanotechnol2008, GershensonJETP1990, BergmannPRB1990}.

In addition to analyzing the applicability of 2TM,  we obtained a number of new results concerning
the energy relaxation of hot electrons. The most important of them are the power dependence of $
\Delta T_e (t) $ in a thermally insulated film [Eq.~(\ref{22})] and the power dependence of $
\Delta T_e (t) $ in thick and thin films at the final stage of relaxation [Eqs.~(\ref{51}) and
(\ref{53})]. The importance of the relation (\ref {22}) is that it allows us to find the parameter
$ r $ from the time dependence $ \Delta T_e (t) $, and this parameter determines the nature of
electron scattering on impurities in the film, as well as the frequency dependence of the
Eliashberg function for the low frequencies.

Interestingly, Eqs.~(\ref{51}) and (\ref{53}) can be used for finding the mean time of elastic
phonon scattering in the substrate $ \tau_{ip} $. It was assumed above that the phonons that "fly"
out to the substrate propagate there ballistically and do not return back to the film. This is
true for narrow films and single-crystal substrates with high thermal conductivity. However, the
return of phonons should be taken into account for non-crystalline substrates with a short mean
free path (for example, glass), as well as wide films, and the transmission coefficient $ \langle
\alpha \rangle \approx 1 $. Returning phonons lead to the dependence $ \Delta T_e (t) \propto
t^{-1/2} $ \cite{Sergeev-SemenovPRB1994} if the time $ t $ is greater than the characteristic
phonon return time $ \tau_{r} \sim \tau_{ip} / \langle \alpha \rangle ^2 $. A similar feature was
observed in experiments on thin niobium films deposited on substrates with low thermal
conductivity \cite{GershensonJETP1990}. Thus, from  the crossover time from the dependence
(\ref{51}) (or (\ref{53})) to the dependence $ \Delta T_e (t) \propto t^{-1/2} $, we can find the
time $ \tau_{r} $, and hence the elastic scattering time of the phonon in the substrate.

 Since our
work uses a  simple model, which, in particular, does not take into account the interaction of
electrons with transverse phonons, it is important to understand how obtained results
 depend on the model. As it follows from the results of Sections~\ref{s3} and \ref{s4}, the
cooling kinetics of heated electrons is determined by the following characteristic times: the
electron-phonon collision time $ \tau_e $, the phonon-electron collision time $ \tau_p $ (which is
$ \nu_{0}^{-1} $ for $ r = -1 $ and is equal to $ \nu(T_B)^{-1} $ for $ r \neq -1 $), as well as
the average time of phonon escape from the film $ \tau_{es} $. Allowance for transverse phonons
will lead to a renormalization of the time $ \tau_e $, since along with longitudinal phonons
electrons can also emit transverse phonons. The time $ \tau_p $ should be replaced by the
phonon-electron collision time, averaged over the phonon modes. In addition, $ \tau_{es} $ should
be understood as the averaged time of departure of longitudinal and transverse phonons from the
film \cite{Little}. In general, the picture of the energy relaxation of electrons with allowance
for transverse phonons will be preserved, and in this respect it does not depend on the model.

It should also be noticed that the formalism presented in  Sec.~\ref{s2}, in principle, makes it
possible to include transverse phonon modes into the model, having formulated the boundary
conditions (\ref{4}) and (\ref{5}) in such a way that they take into account the conversion of
longitudinal and transverse modes of lattice vibrations at the boundaries of the film
\cite{Kaplan}. At the same time, however, the expressions describing the dynamics of the electron
temperature become significantly more complicated and  less transparent.

Since the time-varying electron temperature cannot be measured directly, low-temperature
experiments usually analyze the dynamics of the voltage across the film where a small measuring
current of a fixed value is passed. Due to the temperature dependence of the phonon contribution
to the resistance, the voltage across the film is a function of the electron temperature, and
hence it can serve as a thermometer for the electron temperature.
\cite{BezuglyjShklovskijJETP1997, Bezuglyj&Shklovskij2014}

The examination of the kinetics of electron cooling in metallic films at low temperatures makes it
possible to draw the following conclusions:

1) In a microscopic theory (which is the main one for this paper), electrons are characterized by
the temperature $ T_e $, and phonons are described by a distribution function. The dynamic
equation for $ T_e (t) $ obtained in microscopic theory turns out to be integro-differential
Eq.~(\ref{11}), which makes its analysis difficult and in general requires numerical methods. In a
relatively simple case of weak heating, the linearized equation for $ \Delta T_e (t) $
Eq.~(\ref{12}) has an analytic solution which, at $ r \neq -1 $, gives a power-law relaxation in a
heat-insulated film, and a power-law $ \Delta T_e (t) $ in the final stage of relaxation in the
presence of a heat sink.

2) In the low-temperature region, 2TM quantitatively describes the kinetics of the electron-phonon
system of the metal film only in  case of $ r = -1 $ when the phonon-electron collision frequency
does not depend on the phonon wave vector. This condition is fulfilled when electrons are mainly
scattered on heavy impurity atoms or grain boundaries, which are weakly involved in the lattice
vibration. As can be seen from a comparison with microscopic theory, at $ r \neq -1 $ 2TM does not
give a qualitatively correct (power-law) relaxation character $ T_e (t) $ at the late stages.

3) 2TM ignores a large group of phonons, in which the angles of incidence on the "film-substrate"
interface exceed the angle of total internal reflection ($ \theta >\theta_{cr} $). This
disadvantage is overcome by  3TM, which takes into account both phonons with angles of incidence $
\theta <\theta_{cr} $ and phonons with angles of incidence $ \theta >\theta_{cr} $. 3TM improves
the description of the energy relaxation in a thin-film system (in comparison with 2TM) and at $ r
= -1 $ gives results that are close to the results of the microscopic model.

\section*{Acknowledgments}
We would like to thank Professor V.V.Kruglyak for discussion. Funding from the European Union's
Horizon 2020 research and innovation programme under the Marie Sklodowska-Curie grant agreement
No.644348 (MAGIC) is acknowledged.

\appendix*

\section{ Relaxation of the electron-phonon system in the framework of 2TM}\label{sa}

The two-temperature model introduced in Ref.~\onlinecite{KaganovLifshitzTanatarov} describes the
relaxation of interacting electrons and phonons in terms of their temperatures $ T_e $ and $ T_p
$. At low temperatures, the dynamic equations for these temperatures have a simple form

\begin{equation}\label{a1}
c_e(T_e)\frac{dT_e}{dt}= -\Sigma_5 (T_e^5-T_p^5),
\end{equation}

\begin{equation}\label{a2}
c_p(T_p)\frac{dT_p}{dt}=  \Sigma_5 (T_e^5-T_p^5)-K(T_p^4-T_B^4),
\end{equation}
where the constants $ \Sigma_5 $ and $ K $ describe the heat transfer between electrons and
phonons in the film and between the phonon subsystems of the film and substrate. For weak initial
heating of the electrons, Eqs.~(\ref{a1}), (\ref{a2}) can be linearized in $ \Delta T_e $ and $
\Delta T_p $, i.e. in small deviations of temperatures $ T_e $ and $ T_p $ from the thermostat
temperature $ T_B $. Using the notation $ \tau_e = c_e / (5 T_B^4 \Sigma_5) $ and $ \tau_{es} =
c_p / (4 K T_B^3) $, we arrive at a set of linear equations

\begin{equation}\label{a3}
\frac{d\Delta T_e}{dt}= -\frac{1}{\tau_e} (\Delta T_e - \Delta T_p),
\end{equation}

\begin{equation}\label{a4}
\frac{d\Delta T_p}{dt}=  \frac{c_e}{c_p\tau_e} (\Delta T_e-\Delta T_p)-\frac{1}{\tau_{es}}\Delta
T_p
\end{equation}
with the initial conditions $ \Delta T_e (0) = \Delta T (0) $ and $ \Delta T_p (0) = 0 $. The set
of Eqs.~(\ref{a3}), (\ref{a4}) is solved by the operational method. For the Laplace transforms $
\Delta \tilde T_e (p) $ and $ \Delta \tilde T_p (p) $, we obtain

\begin{equation}\label{a5}
\Delta \tilde T_e(p)=\frac{(p+\nu_p+\nu_{es})\Delta T(0)}{p^2+
(\nu_e+\nu_p+\nu_{es})p+\nu_e\nu_{es}},
\end{equation}

\begin{equation}\label{a6}
\Delta \tilde T_p(p)=\frac{\nu_p\,\Delta T(0)}{p^2+ (\nu_e+\nu_p+\nu_{es})p+\nu_e\nu_{es}},
\end{equation}
where $ \nu_e = 1 / \tau_e $, $ \nu_p = c_e \nu_e / c_p $ and $ \nu_{es} = 1 / \tau_{es} $. We
denote the roots of the common denominator in Eqs.~(\ref{a5}), (\ref{a6}) by $ \kappa_1 $, $
\kappa_2 $. Their values are as follows:

\begin{eqnarray}\label{a7}
&\kappa&_{1,2} =-(1/2)(\nu_e+\nu_p+\nu_{es})
 \pm
(1/2)(\nu_e^2+\nu_p^2+\nu_{es}^2 \nonumber \\
&+& 2\nu_e\nu_p+2\nu_p\nu_{es}- 2\nu_e\nu_{es})^{1/2}.
\end{eqnarray}
It is convenient to write the solution of the set of Eqs.~(\ref{a3}), (\ref{a4}) via $ \kappa_1 $
and $ \kappa_2 $. We have

\begin{equation}\label{a8}
\Delta T_e(t)=\frac{\Delta T(0)}{\tau_e(\kappa_1-\kappa_2)}\Bigl [(\tau_e\kappa_1+1)e^{\kappa_2 t}
-  (\tau_e\kappa_2+1)e^{\kappa_1 t}\Bigr ],
\end{equation}

\begin{equation}\label{a9}
\Delta T_p(t)=\frac{\Delta
T(0)}{\tau_e(\kappa_1-\kappa_2)}(\tau_e\kappa_1+1)(\tau_e\kappa_2+1)\Bigl  (e^{\kappa_2 t} -
e^{\kappa_1 t}\Bigr ).
\end{equation}

In the main part of the paper we considered the case of a poor thermal bond between a film and a
substrate, when the strong inequality $ \nu_e, \nu_p \gg \nu_{es} $ holds. In this case, 2TM gives
for the electron temperature the expression

\begin{equation}\label{a10}
\Delta T_e(t)=\frac{c_p \Delta T(0)}{c_p+c_e}e^{-(\nu_e+\nu_p) t} +  \frac{c_e \Delta
T(0)}{c_p+c_e}e^{-\nu_{es} t/(1+\zeta)}.
\end{equation}
Here we use the relation $ c_e \nu_e = c_p \nu_p $ and the notation $ \zeta = c_e / c_p $.

\end{document}